\documentclass[conference]{IEEEtran}
\IEEEoverridecommandlockouts
\usepackage{cite}
\usepackage{amsmath,amssymb,amsfonts}
\usepackage{algorithmic}
\usepackage{graphicx}
\usepackage{textcomp}
\usepackage{xcolor}

\usepackage[colorlinks=true, allcolors=blue]{hyperref}
\usepackage{todonotes}
\usepackage{listings}
\usepackage{csquotes}

\lstdefinestyle{pythonstyle}{
    language=Python,
    basicstyle=\ttfamily\small,
    keywordstyle=\color{blue}\bfseries,
    stringstyle=\color{green!70!black},
    commentstyle=\color{gray},
    numberstyle=\tiny\color{gray},
    numbersep=5pt,
    showstringspaces=false,
    breaklines=true,
    frame=single,
    tabsize=4
}
\usepackage{enumitem}

\newcounter{virgilecounter}

\newcounter{yanniscounter}

\newcounter{paulcounter}

\def\BibTeX{{\rm B\kern-.05em{\sc i\kern-.025em b}\kern-.08em
    T\kern-.1667em\lower.7ex\hbox{E}\kern-.125emX}}
    
\begin{document}

\title{ReservoirChat: Interactive Documentation Enhanced with LLM and Knowledge Graph for ReservoirPy
\thanks{* These authors are co-first authors. $\dagger$ Corresponding author.
}
%\thanks{}
}

\author{\IEEEauthorblockN{Virgile Boraud*}
\IEEEauthorblockA{\textit{Mnemosyne} \\
\textit{Inria}\\
Bordeaux, France \\
virgile.boraud@yahoo.fr}
\and
\IEEEauthorblockN{Yannis Bendi-Ouis*}
\IEEEauthorblockA{\textit{Mnemosyne} \\
\textit{Inria}\\
Bordeaux, France \\
yannis.bendi-ouis@inria.fr}
\and
\IEEEauthorblockN{Paul Bernard}
\IEEEauthorblockA{\textit{Mnemosyne} \\
\textit{Inria}\\
Bordeaux, France \\
paul.bernard@inria.fr}
\and
\IEEEauthorblockN{Xavier Hinaut$\dagger$}
\IEEEauthorblockA{\textit{Mnemosyne} \\
\textit{Inria}\\
Bordeaux, France \\
xavier.hinaut@inria.fr}
}

\maketitle

\begin{abstract}
%%Various methods have been explored to improve Large Language Models (LLMs) for enhancing coding assistance with ReservoirPy and providing accurate answers to complex questions in the field of Reservoir Computing. By leveraging external knowledge through Retrieval-Augmented Generation (RAG) and knowledge graphs, the proposed tool aims to reduce hallucinations and improve the accuracy of LLM-generated responses. Our goal is to create a platform similar to ChatGPT, specifically designed to help users write and correct Python code using the ReservoirPy library, while being capable of reliably answering complex technical questions regarding Reservoir Computing. After testing, although the most recent but private solutions like ChatGPT-4o or NotebookLM are still slightly better for knowledge-related questions, our solution outperforms them on code-related questions and significantly outperforms its base model, Codestral-22B.

We introduce a tool designed to improve the capabilities of Large Language Models (LLMs) in assisting with code development using the ReservoirPy library, as well as in answering complex questions in the field of Reservoir Computing. By incorporating external knowledge through Retrieval-Augmented Generation (RAG) and knowledge graphs, our approach aims to reduce hallucinations and increase the factual accuracy of generated responses. The system provides an interactive experience similar to ChatGPT, tailored specifically for ReservoirPy, enabling users to write, debug, and understand Python code while accessing reliable domain-specific insights. In our evaluation, while proprietary models such as ChatGPT-4o and NotebookLM performed slightly better on general knowledge questions, our model outperformed them on coding tasks and showed a significant improvement over its base model, Codestral-22B.
\end{abstract}


%% NOTES A GARDER
% Les QA sont sur cette page : https://notes.inria.fr/xki6i3DGSs6ItPNJZSt6Ow#

\section{Introduction}

Large Language Models (LLMs) are transformer-based models \cite{vaswani2017attention} designed to generate text, that users interact with using natural language. These models are pre-trained on a large quantity of data covering a variety of domains and linguistic styles \cite{achiam2023gpt}. Training data is usually imported from web pages, newspaper articles, books, documentation or literature \cite{radford2018improving} \cite{radford2019language} \cite{brown2020language}. 
Llama3 \cite{dubey2024llama} was trained on 15 trillion \( (1.5 \times 10^{13}) \) tokens. However, training on such large datasets requires significant computational resources, raising concerns about time, cost, and energy consumption \cite{singh2024survey}.

Training these models on large code datasets enables them to assist users and developers, showing strong potential for code generation. A notable example is \emph{GitHub Copilot} \cite{peng2023impact}, an assistant designed to provide suggestions, error corrections, and autocompletions of code. It is powered by OpenAI’s Codex model \cite{chen2021evaluating}, a general purpose programming model. Copilot analyzes the context in which code is written, and generates contextually relevant suggestions in real-time, potentially accelerating the coding process.

However, one major problem with LLMs is when the model generates incorrect or unfounded information. This phenomenon is called hallucination \cite{huang2023survey}. It is a particularly concerning problem in specialized or underrepresented domains in the training dataset, like Reservoir Computing \cite{jaeger2001echo}. As such, even if ChatGPT-4o \cite{hurst2024gpt} knows what a reservoir is, it could first confuse it with a water reservoir. Even when clarified within the context of machine learning, it will not always give an accurate answer, and it could provide answers that seems plausible but are actually incorrect.

Reservoir Computing models are not mainstream, despite their performance in different tasks, particularly those involving small sequential datasets or challenging corpora with complex non-linear variable interactions such as COVID hospitalization predictions \cite{ferte:hal-04693930}. This under-representation can make it difficult for users, especially those new to the field, to fully appreciate and harness the potential of Reservoir Computing, and in fact to efficiently parametrize reservoir models.

\emph{ReservoirPy}~\cite{trouvain2020reservoirpy,trouvain2022create,trouvainHALjmlr} %,} 
is a Python library that allows users to easily create and manipulate Reservoir Computing models. But, ReservoirPy users can face challenges when trying to configure reservoir models to understand the underlying intuitions and correctly parametrize the models. Moreover, reading the full documentation can be time-consuming, especially for new users. An interactive assistant may offer a more intuitive and accessible alternative to static documentation, helping users efficiently locate specific features without getting lost in the API.
More generally, Reservoir Computing needs some expertise to properly set up a  model. For instance, hyperparameter tuning could be a challenge for some tasks\footnote{For instance, there is the misconception that values above 1 should not be explored for the \emph{spectral radius} because of theoretical insights that are not well understood on the echo state property.}, and ReservoirPy propose hyperparameter tutorials~\cite{hinaut2021hype}. 

In the following, we propose various versions of an improved LLM augmented with code and documents. Then, we compare its performance with its base model and other mainstream models. Finally, we discuss the strengths and limitations of this approach.

\section{Methods}

\subsection{Embeddings and Simple ChatBot}

ReservoirChat development began with a simple chatbot designed to answer 245 predefined questions (see GitHub repo), with their corresponding answers. The goal was to determine if a simple application, without the involvement of text generator (LLM decoder), could be sufficient. The entire database was converted into vectors using text embeddings, in our case according to the \textit{nomic-embed-text-v1.5} \cite{nussbaum2024nomic} model (LLM encoder) from Nomic-AI with \emph{LM Studio}. Each user-submitted question was vectorized and compared with the database through a cosine similarity process, in order to make correlations between the questions and the database. Based on the closest match, a response would be selected and delivered to the user. This basic approach becomes more effective as the database grows.

When a user submitted a query, the chatbot retrieved the top five most similar embeddings based on cosine similarity between the query embedding and the database of prepared questions. 

The system's limitations became apparent when users asked questions that were not explicitly covered in the database. This issue was exacerbated when handling coding-related questions, where even a slight variation in the code or phrasing of the question could result in the chatbot failing to find a suitable match. This revealed the need for a more sophisticated system capable of handling a wider range of queries, including those that required deeper contextual understanding or more complex reasoning challenges. This preliminary model served as a baseline, limited to simple, predictable questions.

\subsection{Retrieval-Augmented Generation and ReservoirRag}

A more elaborate approach consists in enhancing an LLM with a database. As stated previously, hallucination occurs when LLMs generate incorrect, but sometimes plausible answers, due to a lack of factual basis. Retrieval-Augmented Generation (RAG) \cite{NEURIPS2020_6b493230} \cite{gao2023retrieval} can mitigate this issue without requiring additional model training by dynamically retrieving relevant information from external documentation sources before generating a response, provided the retrieved data has been properly validated. However, if incorrect sources are introduced, RAG will propagate these errors, hence making careful conservation of the knowledge base essential. Assuming the database remains up-to-date, RAG can be easily adjusted with new information at a significantly lower computational cost compared to retraining an entire model. Since RAG influences the accuracy and reliability of responses, selecting an appropriate LLM remains a critical decision, as some models offer superior multilingual capabilities while others excel in coding assistance.

RAG requires the database to be prepared and cut into chunks. These chunks of text provide a basis for generating responses, treated as text embeddings, and correlated according to the previous model. Chunk size is defined by the user depending on the context and parameters. It is possible, for example, to choose segments of 10 lines, entire paragraphs, or even complete portions of code. In addition, it is sometimes appropriate to overlap chunks to maintain context coherence.

Thus, the chatbot evolved into ReservoirRAG, when we decided to connect Codestral-22B \cite{mistral2024codestral} to a RAG. Codestral is a multilingual model developed by Mistral, known for its effective handling of coding queries. The base data consists of documents, including the ReservoirPy documentation, relevant papers about Reservoir Computing, and code samples. After processing, embeddings are generated from these documents and stored in a CSV file alongside their content. These precomputed embeddings are then loaded using the \emph{pandas} library and organized into a DataFrame. Some additional documents were processed in the same manner, but reserved as external data for RAG.

Again, embedding relevance was based on the same test as before, with a 0.75 correlation threshold. This implementation ensures that only the most relevant data is considered, enhancing response accuracy. Once the relevant documents are identified, they are incorporated into the LLM’s response generation process as part of the contextual information, alongside the conversation history, enabling the model to retain memory of previous queries and responses. For queries related to coding, the additional custom document codes.md, is added to the database, containing various code snippets and samples from the ReservoirPy GitHub. This feature is particularly important for coding-related queries, as even minor errors in generated code could lead to incorrect answers.

\subsection{Knowledge graph, GraphRAG and ReservoirChat}

\begin{figure}[h]
\caption{Clustered Graph \enquote{Layer} 0 obtained for ReservoirChat$_{Little}$}
\includegraphics[scale=0.4]{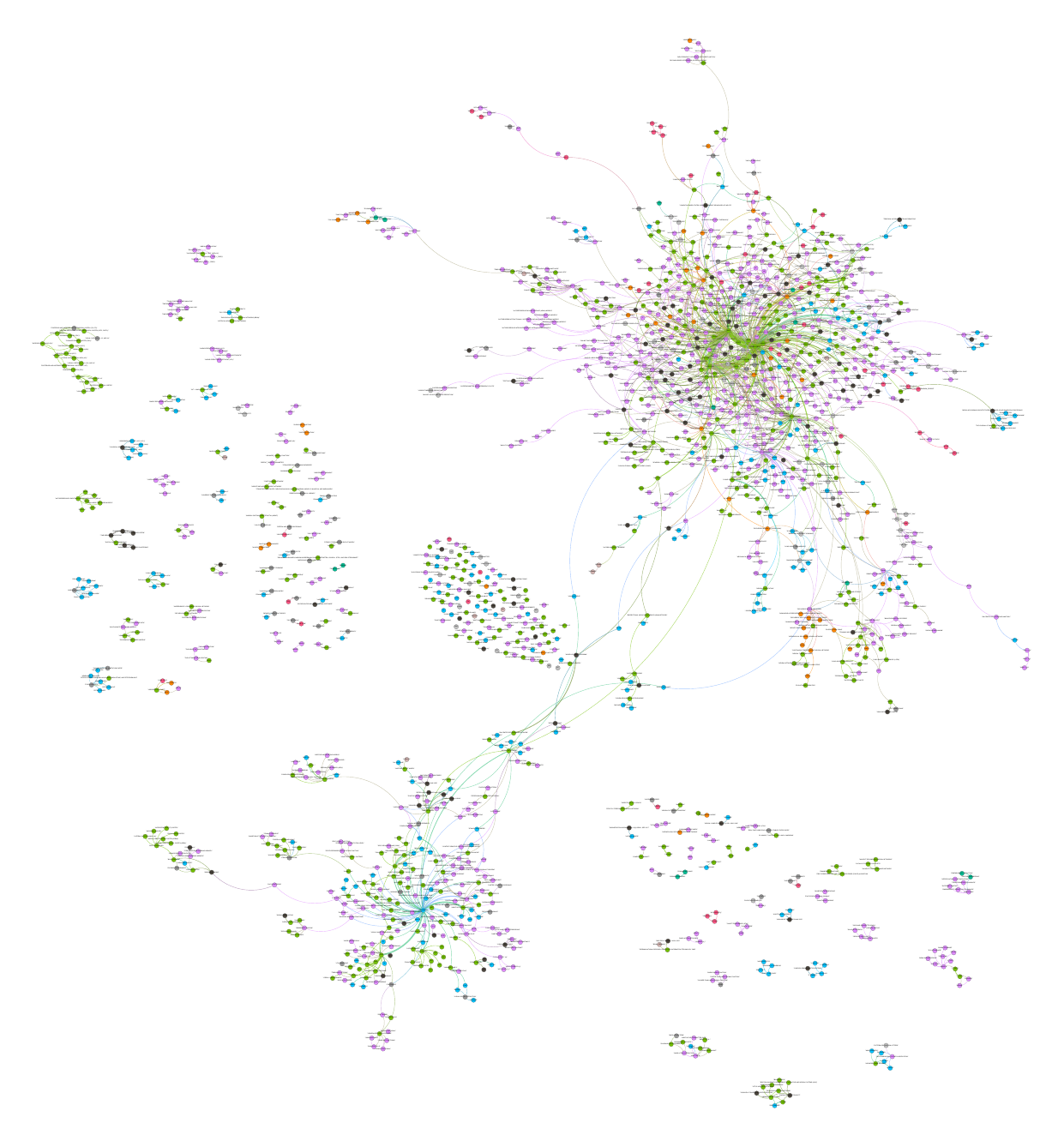}
\centering
\label{fig-clustered-graph-reservoirchat}
\end{figure}

A Knowledge Graph (Figure \ref{fig-clustered-graph-reservoirchat}) \cite{chen2020knowledge} is a structured representation of a dataset that organizes information into entities and their relationships, represented respectively as nodes and edges. Knowledge graphs allow LLMs to associate different concepts in a more qualitative manner, improving the contextual and semantic accuracy of answers. Unlike simply implementing RAG, this method requires preprocessing the documents during the initial creation of a graph, which can become time and resource intensive for a complete and thorough result. It is nevertheless lighter than retraining the entire model.

\begin{figure}[h]
\caption{Zoom of the graph. We can see the links between reservoirpy and other entities}
\vspace{10pt} 
\includegraphics[scale=2.6]{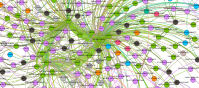}
\centering
\label{fig-clustered-graph-reservoirchat-zoom}
\end{figure}

The GraphRAG \cite{edgetrinh2024graphrag} method, developed by Microsoft, stands out for its ability to handle a large and complex volume of data, in the form of corpora, such as scientific papers. Where traditional RAG shines when it comes to specific questions, GraphRAG is more versatile, and if effective at answering both specific questions, and broader queries about the general theme of a corpus.

\begin{figure}[h]
\caption{Pipeline Stage: Construction of the Knowledge Graph. Image from \cite{edgetrinh2024graphrag}.}
\includegraphics[scale=0.45]{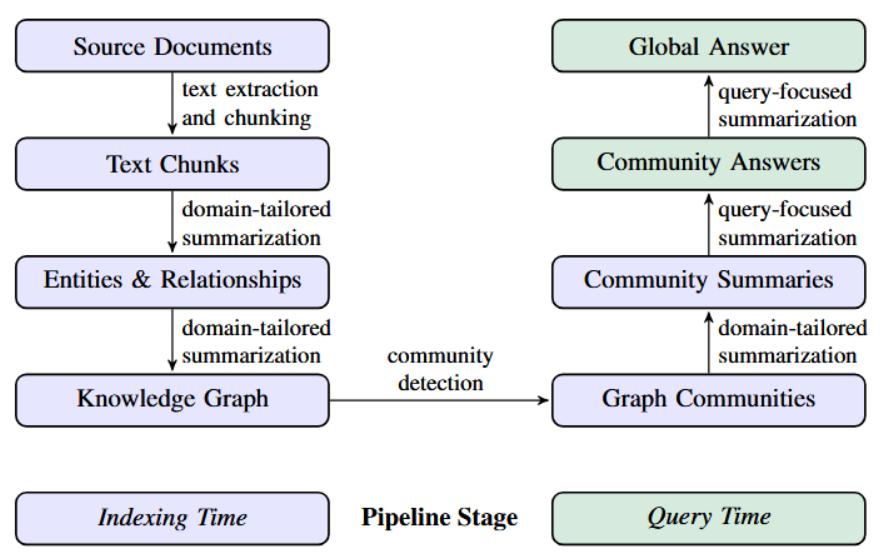}
\centering
\label{fig-pipeline}
\end{figure}

According to the methodology (Figure \ref{fig-pipeline}) described by \cite{edgetrinh2024graphrag}, the knowledge graph is constructed starting from a database of documents, which are cut into chunks as detailed before. Using LLM-driven prompts, the system identifies named entities and the relationships between them. This is repeated in several rounds to ensure that all relevant entities are captured. Each individual element is summarized according to a domain-tailored summarizing process. Then, the system builds an undirected graph in which each node represents an entity, and each edge captures the relationship between two entities. The edges are weighted based on the frequency and strength of the relationships. Algorithms like Leiden \cite{Traag_2019} are used to partition the newly created graph, detecting clusters of closely related nodes within the graph. These clusters, also known as communities, reflect topics or themes that are highly interconnected (Figure \ref{fig-clustered-graph-reservoirchat-zoom}). A summary is later generated for each community, helping to create partial answers to user queries by drawing from multiple related entities. These partial or local answers may be relevant on their own, or summarized even further into a so-called global answer. In the case of ReservoirChat and its final version, it was more appropriate to mainly utilize local answers, as they provide more focused and detailed information than a global response. While more comprehensive, global responses tend to be less precise, which is intuitively less desirable for the scientific and educational objectives of ReservoirChat.

\begin{figure}[h]
\caption{Comparison of training documents used for each ReservoirChat and NotebookLM versions}
\includegraphics[scale=0.55]{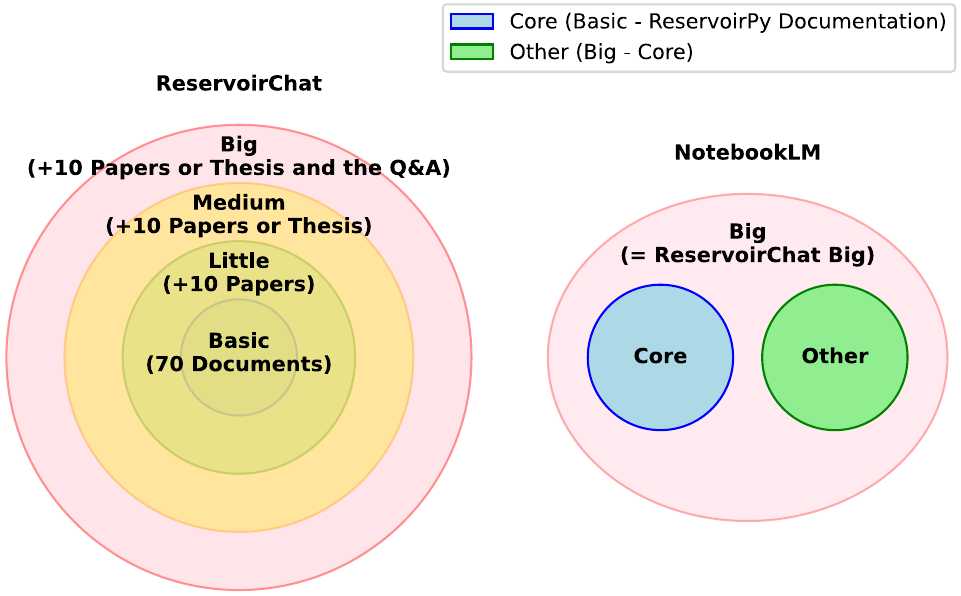}
\centering
\label{fig-comparison-training-doc}
\end{figure}

Several versions of ReservoirChat were created with varying content and database sizes (Figure \ref{fig-comparison-training-doc}). The Basic version includes resolved issues related to ReservoirPy, its complete documentation and code, along with code samples, resulting in a corpus of 70 documents. The Little version includes 10 additional research papers about Reservoir Computing. Medium has 10 more research documents appended to its database, including a PhD thesis. Finally, Big includes all previous resources with 10 additional scientific papers, and a compilation of 245 prepared questions and answers, including 55 code-related questions, and 195 about various concepts related to Reservoir Computing and ReservoirPy. The Big version is the one currently available on the website.

However, GraphRAG also has its limitations. The construction of the permanent knowledge graph requires the precise extraction of entities and relations from the source documents. This requires additional data preprocessing to guarantee the quality of the graph. It also needs to be regularly updated with new information to stay relevant, which can be computationally expensive.

\subsection{Model and web interface implementations}

\begin{figure}[h]
\caption{ReservoirChat$_{Big}$ Interface}
\includegraphics[scale=0.3]{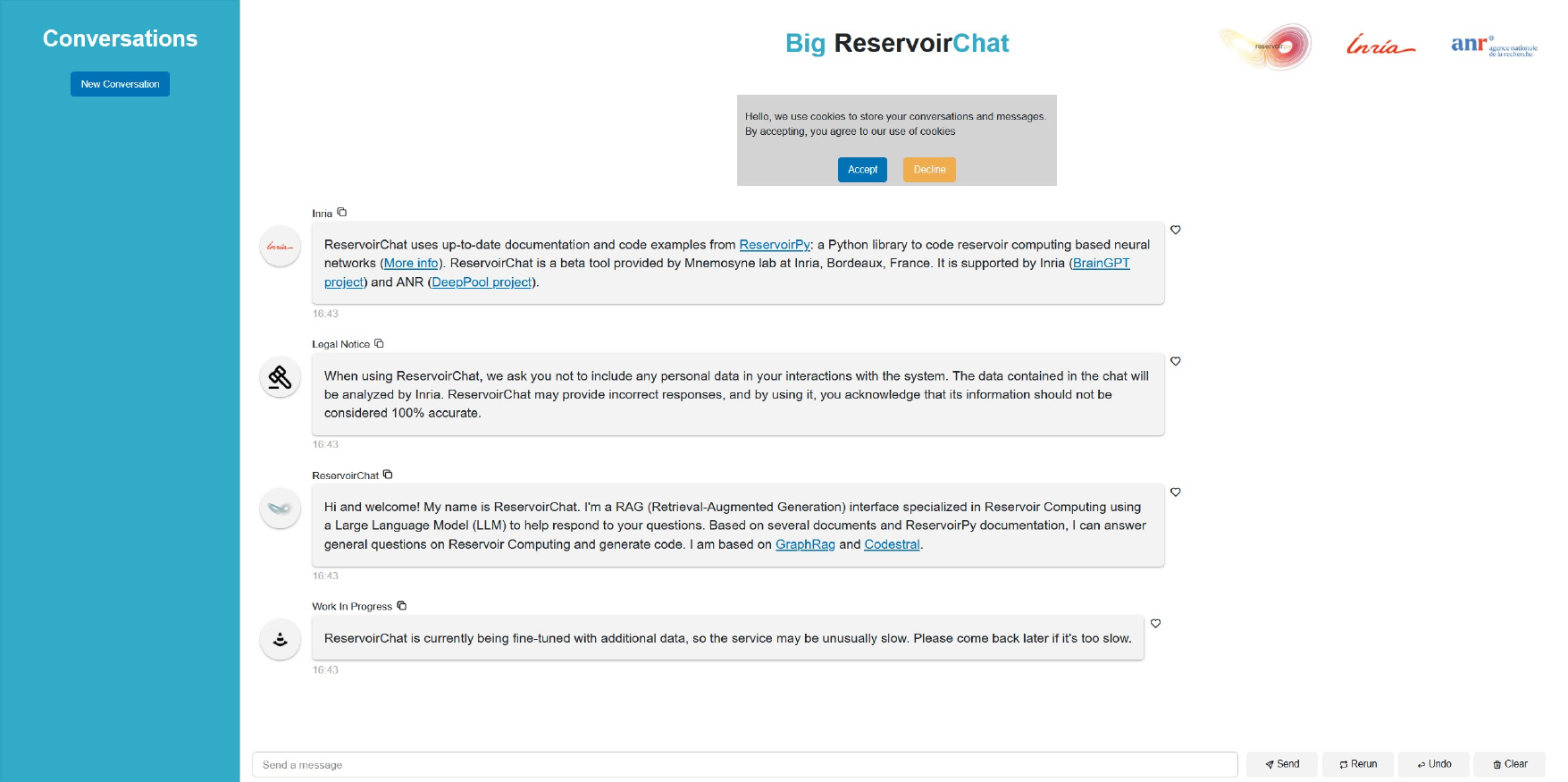}
\centering
\label{fig-reservoirchat-interface}
\end{figure}
The user interface (Figure \ref{fig-reservoirchat-interface}) can be accessed at  https://chat.reservoirpy.inria.fr/). 
In addition, code and most documents to create the various versions of ReservoirChat models can be found in the following GitHub repository: \url{https://github.com/Naowak/ReservoirChat/}.
Some documents that can be freely accessed elsewhere (such as scientific publications content) were not included.
An Appendix with benchmark questions is also available at the same URL.

\subsection{Evaluation method}

Response quality in natural language is complex to evaluate, hence the need to consider a systematic approach. Human evaluation is not preferred, as it would both be highly time-consuming and require expertise, which is costly, in addition to being vulnerable to subjectivity and cognitive bias. LLM-based evaluation was also excluded, as there are currently no LLM specifically trained about ReservoirPy and Reservoir Computing, therefore leading to potential imprecision and unreliability.

Benchmarking was the chosen method, being the most objective one. A set of questions, written by a Reservoir Computing expert, were given to the LLM. The model was tasked with selecting the correct answer from 4 suggestions. 

In the following experiment, we compared every model of ReservoirChat we created with ChatGPT-4o, Llama3, Codestral and NotebookLM \cite{google2024notebooklm} (based on Gemini 1.5 \cite{team2024gemini}). Only ReservoirChat and NotebookLM were fed with documents about Reservoir Computing and ReservoirPy. ChatGPT-4o, Llama3 and Codestral relied only on their own knowledge and training methods to answer. We tried different sets of documents for NotebookLM. NotebookLM$_{Core}$ includes the same documents as ReservoirChat$_{Basic}$ except the documentation. NotebookLM$_{Other}$ includes every document not included in the core version, this was added to test the influence of the core model compared to the other model. The NotebookLM$_{Big}$ represents all the documents like ReservoirChat$_{Big}$. For every model a temperature of 0.1 was used to minimize stochastic effects, as preliminary tests showed it was a good compromise and close to a temperature of 0 \footnote{Temperature in an LLM controls freedom in responses, on a scale from 0 to 1, 0 resulting in more deterministic answers with no variation. This setting influences the model's creativity in generating answers.} which is often preferred \cite{renze2024effect}. However, ChatGPT-4o  does not offer the possibility to change temperature, nor does NotebookLM. 

The benchmark consisted of 20 knowledge-based questions and 14 coding-related questions, including 8 debugging tasks. To minimize bias, each question was asked independently, without prior context. A standardized prompt designed to reduce any disparity was provided to ensure fair comparison across all models : 

\blockquote{You will be given a question with four answer options labeled A, B, C, and D. Please respond using the library ReservoirPy with only the letter (A, B, C, or D) that is the correct answer.}

Each question were asked 3 times, therefore a score of 0, 0.33, 0.66, or 1 respectively indicates that the model gave the correct answer zero, once, twice or three times out of three attempts. The total score of a model is the addition of the individual score for each question, respectively on a scale from 0 to 20 for the first benchmark, then from 0 to 14 for the second, with 0 indicating the lowest possible performance.

\section{Results}

\subsection{Brief Recapitulation}

\begin{figure}[h]
\centering
\caption{Results of knowledge questions from the 20 questions benchmark for ChatGPT 4o (GPT 4o), LLama3, Codestral, NotebookLM (NLM) and ReservoirChat (RC). The source model (Codestral) and ReservoirChat models is framed in black.}
\vspace{0.3 cm}
\includegraphics[scale=0.5]{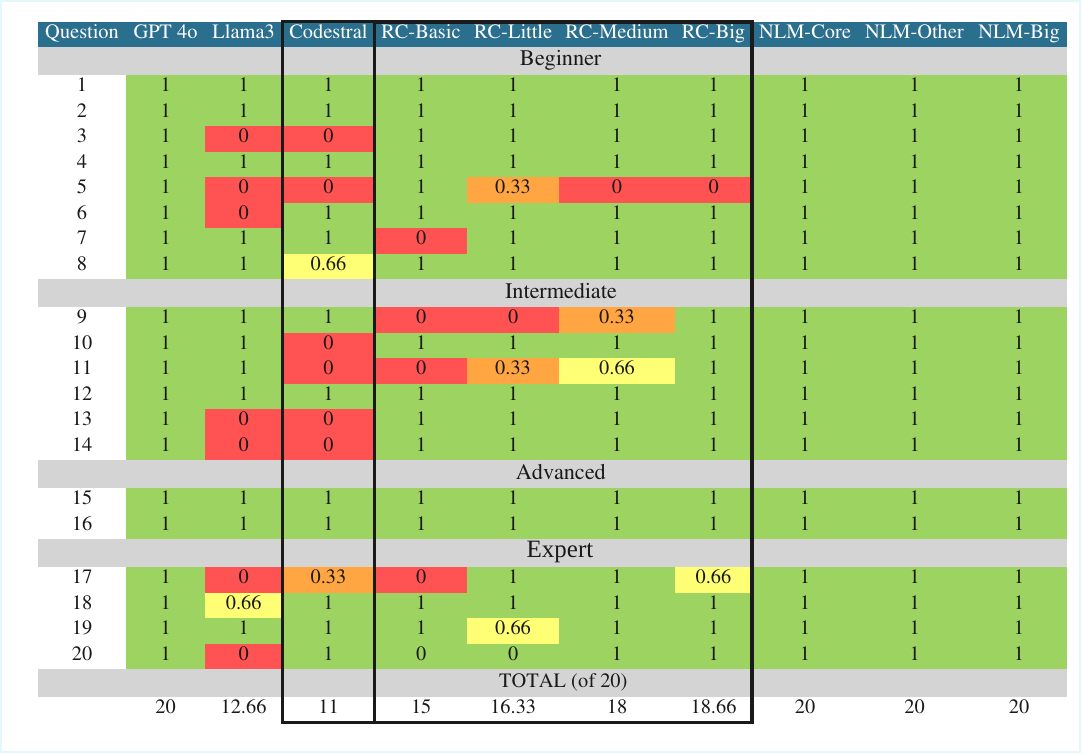}
\label{fig-result-knowledge}
\end{figure}

\begin{figure}[h]
\caption{Results of code and debug questions from the 14 questions benchmark for ChatGPT 4o (GPT 4o), LLama3, Codestral, NotebookLM (NLM) and ReservoirChat (RC).}
\vspace{0.3 cm}
\includegraphics[scale=0.5]{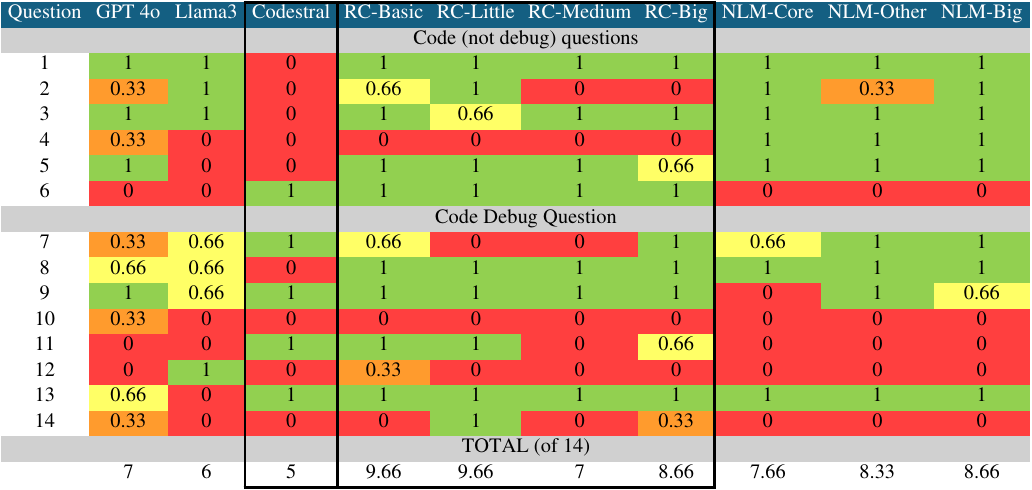}
\centering
\label{fig-result-code}
\end{figure}

In this section, we aim to compare ReservoirChat, which is based on the Codestral architecture, to various classical models such as LLaMA3 and GPT-4o, as well as other models capable of learning from external databases, such as NotebookLM. Specifically, the results focus on evaluating two important aspects: knowledge-based questions related to Reservoir Computing concepts and coding/debugging tasks involving the ReservoirPy library. We also examine how incorporating external resources, such as scientific papers and documentation, impacts the performance of several models. The evaluation uses a custom-made benchmark developed to assess the accuracy and consistency of responses in the context of Reservoir Computing (see Methods). 

\subsection{Analysis of Knowledge-Based Performance}

Regarding knowledge questions (Figure \ref{fig-result-knowledge}), Llama3 and Codestral have comparable score of 12.66 and 11, respectively. All ReservoirChat models output higher scores than the default Codestral, increasing with each addition of documents from a score of 15 for the Basic model, up to a score of 18.66 with the ReservoirChat$_{Big}$. Performance improves proportionally with the number of documents added to ReservoirChat’s database. Further improvements could be achieved by adding more documents and fine-tuning the model. Also ChatGPT-4o and all NotebookLM versions obtain a score of 20. This perfect score highlights a superior ability to answer knowledge-based questions compared to other models.

\subsection{Analysis of Code and Debug Performance}

On the other hand, code and debug questions (Figure \ref{fig-result-code}) have an overall much lower performance for all models, with none having or approaching a perfect score. ChatGPT-4o performed moderately at 7. NotebookLM ranges from 7.66 to 8.66. It is interesting to note that the Core version, trained on resolved issues, has a worse score than the Other version which includes ReservoirPy documentation and scientific papers about reservoirs and the library. The difference, slightly in favor of NotebookLM$_{Other}$, equals to 8.33 - 7.66 = 0.67, or 4.8\%, could indicate that an LLM trained on a whole language documentation is more reliable in coding assistance than one only trained on resolved issues. NotebookLM$_{Big}$ outputs a similar score of 8.66. 
These results may be biased, as in real-world scenarios, developers often face novel issues not covered by existing resolved cases, potentially lowering NotebookLM$_{Core}$'s performance. 

Llama3 yields better results than Codestral, with a score of 6 against 5. Given that Codestral is known for its reliability in coding assistance, this result surprising, and suggest that as ReservoirPy is a recent library, the model may not have been trained on a lot of data apart from the documentation in itself. Codestral's usual reliability must then be founded on training with larger or different type of dataset. This discrepancy may also be due to how the questions were posed, or the fact that Codestral’s training corpus may focus less on English and more on other languages.

ReservoirChat models also outperform Codestral, obtaining a score of 9.66 for the Basic and Little Models, 7 for Medium, and 8.66 for Big. These results may be indicative of the influence of other non-code related documents on the Knowledge Graph. The larger the database, the more complex and numerous the connections between entities are, potentially affecting some answers. Also, Medium having the lowest score for coding questions, out of all the ReservoirChat versions tested, could indicate that there is a limit up to which more data confuses the model but there is not enough data to adjust its own performance yet.

Finally, it is interesting to note when asked three times, ChatGPT-4o's provided different answers to 50\% of the coding questions, suggesting that it has a high temperature. In comparison, when asked three times, ReservoirChat$_{Big}$ provided different answers to only 21.4\% of the coding questions. Also, questions were arranged into categories and levels of difficulty, but no evident differences were observed.

\begin{table*}[h]
\caption{Score difference table of every ReservoirChat models compared to Codestral, NotebookLM$_{Core}$, NotebookLM$_{Other}$ and NotebookLM$_{Big}$.
How to read: given an existing solution (other models in columns), how much do we gain/loose  (in score points) if we change for our solution (our models in lines). 
Formula: our-model[lines] - other-models[columns].}
\includegraphics[scale=0.63]{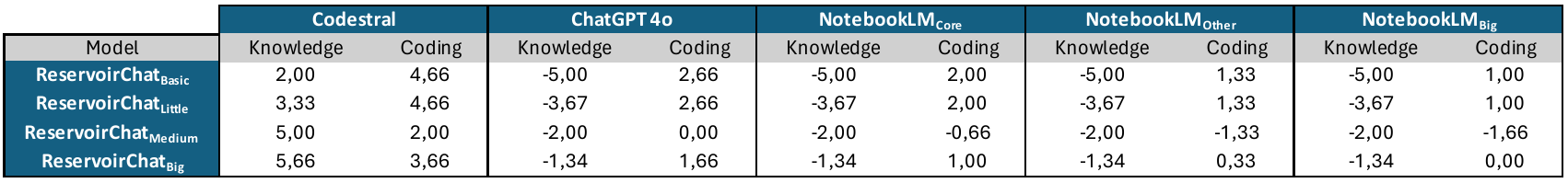} 
\centering
\label{tab-difference-table}
\end{table*}

\begin{table*}[h]
\caption{Performance Table of every ReservoirChat models compared to Codestral, NotebookLM$_{Core}$, NotebookLM$_{Other}$ and NotebookLM$_{Big}$.
How to read: given an existing solution (other models in columns), how much do we gain/loose (in percentage) if we change for our solution (our models in lines). 
Formula: 100 * (our-model[lines]  - other-models[columns]) / other-models[columns].
}

\includegraphics[scale=0.63]{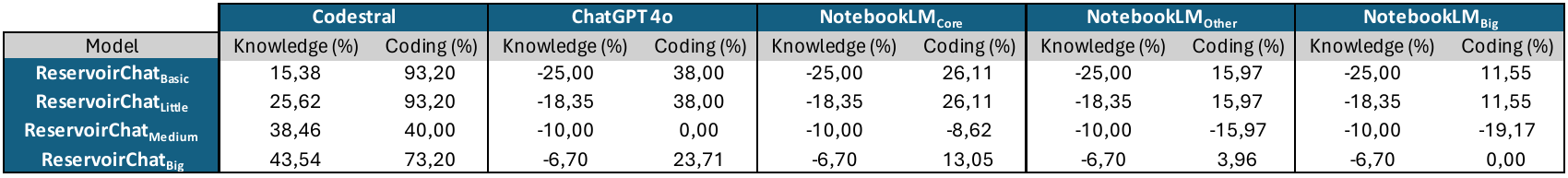}
\centering
\label{tab-difference-table-percentage}
\end{table*}

ReservoirChat${Big}$ is slightly outperformed (Table \ref{tab-difference-table} and \ref{tab-difference-table-percentage}) by ChatGPT 4o and NotebookLM${Big}$ in the case of knowledge questions, with a 6.70\% lower score in both cases. However, its coding score remains comparable to NotebookLM${Big}$, with no difference, and is better by 23.71\% in favor of ReservoirChat${Big}$ in the case of ChatGPT 4o. In other cases, ReservoirChat${Big}$ significantly outperforms other major models, achieving up to a 43.54\% increase in knowledge scores and a 73.20\% improvement in coding compared to Codestral. These results show that even if ReservoirChat doesn't achieve a perfect score, it clearly improves upon Codestral when responding to Reservoir specific questions. These performance differences were calculated as percentage rates based on the total score of each model compared to each ReservoirChat models.

\subsection{Correlation Analysis Across Models}

\begin{figure}[h]
\caption{Pearson correlation matrix of every model for the knowledge related questions}
\includegraphics[scale=0.45]{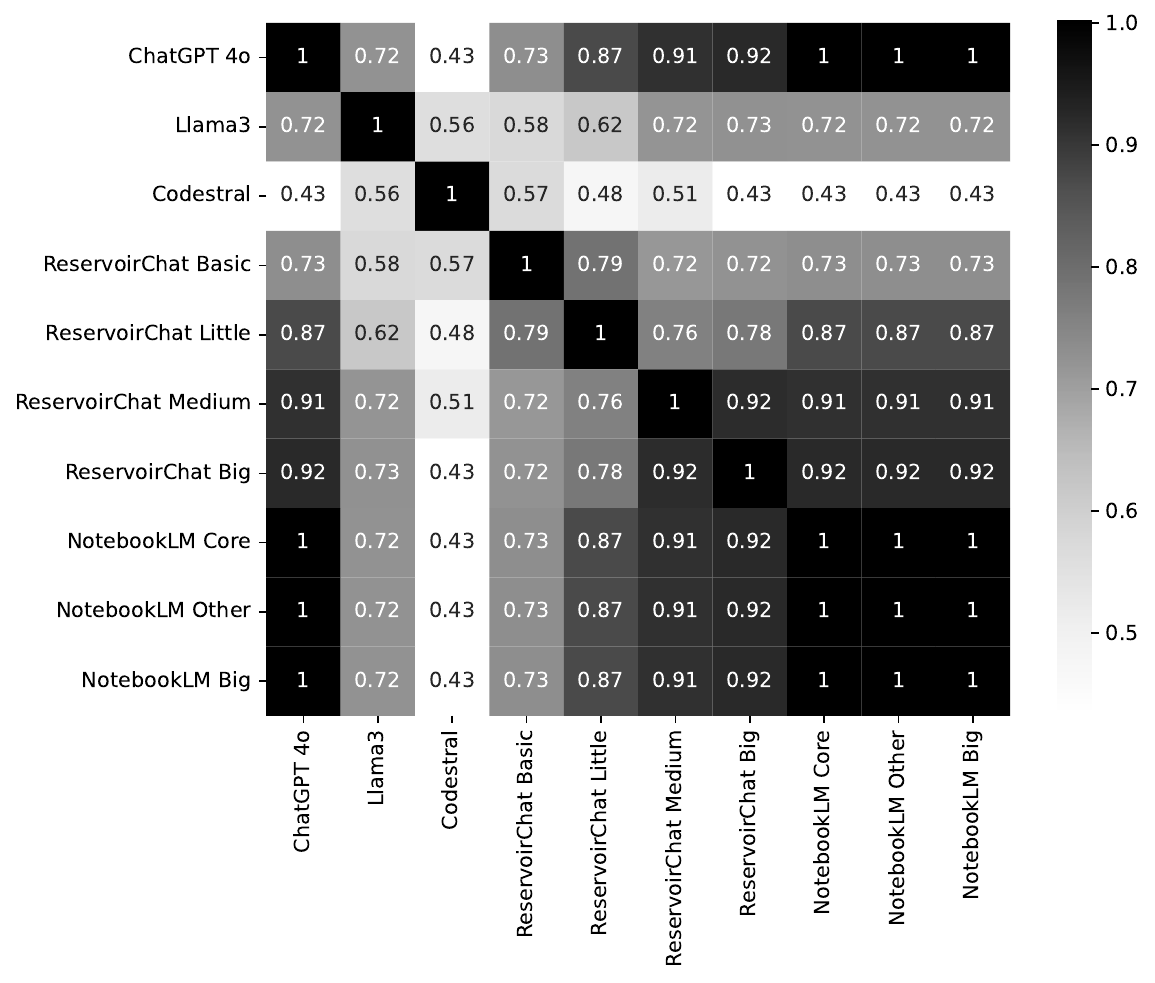}
\centering
\label{fig-correlation-knowledge}
\end{figure}

% \vspace{-1cm}

\begin{figure}[h]
\caption{Pearson correlation matrix of every model for the code related questions}
\includegraphics[scale=0.45]{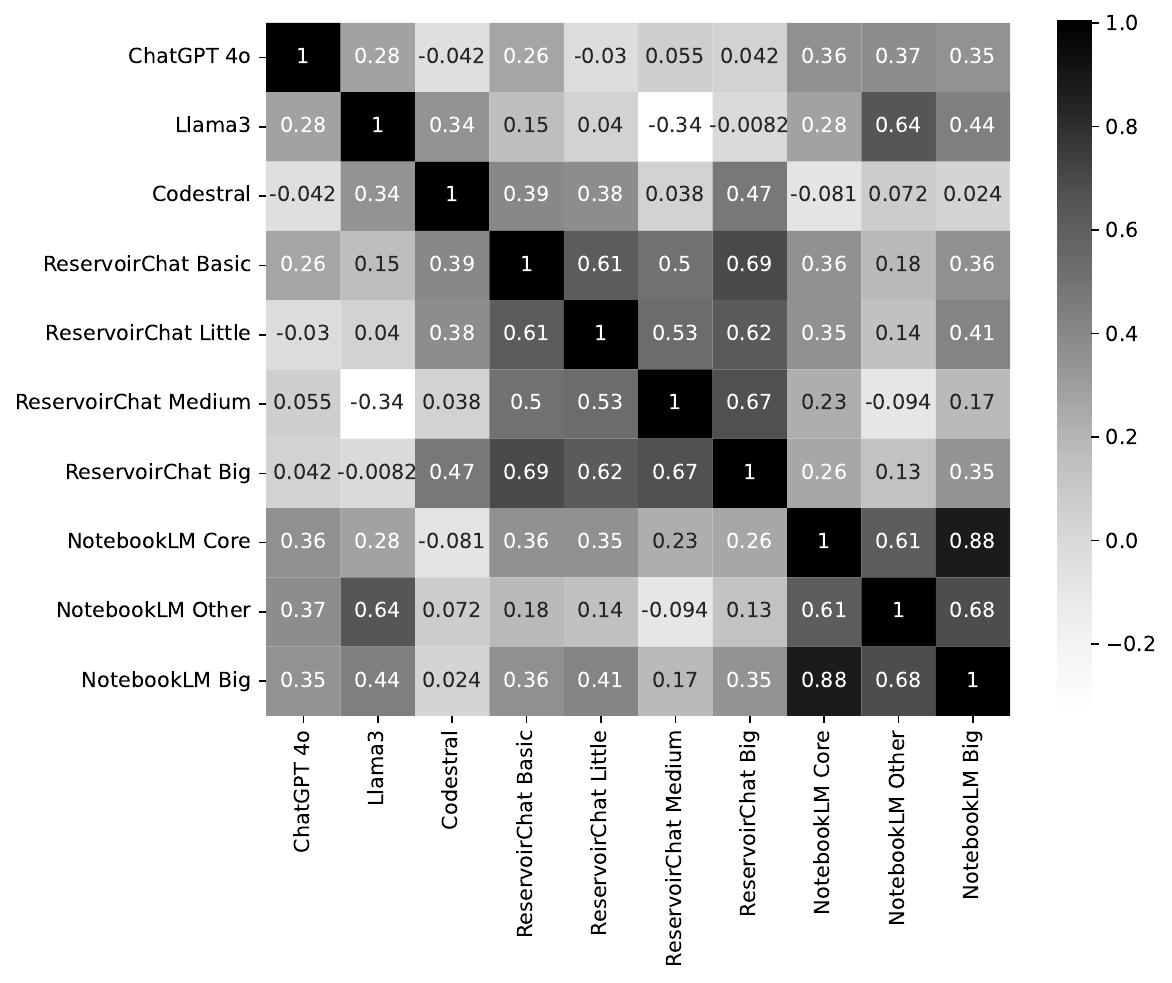}
\centering
\label{fig-correlation-code}
\end{figure}

For each model, we respectively concatenated the 3 answers they gave for the knowledge-related and the code-related questions into two vectors, and we computed the Pearson correlation between all of them. We can see those results in the Figure \ref{fig-correlation-knowledge} and \ref{fig-correlation-code} that respectively show the knowledge-related and code-related correlation matrices. You can find the correlation matrix on the whole questions in the Appendix (available on our repository).

As we can expect, models that perform well are correlated together, because there is only one good answer per question. So, for the knowledge part where lots of models obtained good results, it is not a surprise to see that a lot of models are well correlated together, as we can see with GPT-4o, and all NotebookLM versions which have a correlation score of 1, as they all answered perfectly on all the knowledge questions. Conversely, the less a model performs, the less it will be correlated with the others, as we can see with Codestral. One exception is Llama3, which does not perform well on the knowledge related questions, but is a bit correlated with other high-performing models.

On the other hand, when we look at the code-related matrix correlation, we can observe two major groups of models : those related to NotebookLM and those related to ReservoirChat. In both groups, models are well correlated together, which is expected, because they respectively derive from the same base. We can also observe this with Codestral that is a bit correlated to ReservoirChat but not at all to other models. In fact, Codestral appears a bit isolated, with correlation score close to 0 with every model except ReservoirChat and Llama3. This might be due to the fact that Codestral is the only model specifically trained for coding tasks. As an unexpected results, we also observe a correlation between Llama3 and NotebookLM Other which might be due to the documents there were fed with, and the near-zero correlation between ReservoirChat${Medium}$ and Codestral. 

\subsection{Summary of Findings and Future Directions}

These results are encouraging for RAG based knowledge graph methods. Despite the base model's limitations, our approach improves its performance in domain-specific tasks, particularly in Reservoir Computing.

\section{User data collection with Big ReservoirChat}

Over three months, 130 users interacted with ReservoirChat${Big}$. Some questions were directly related to the ReservoirPy library, while others required broader coding knowledge beyond the preloaded database, relying on Codestral’s base model. 
On average, people asked between two to five questions, indicating that while the tool raises curiosity, it’s not yet fully utilized as an assistant.

We analyzed data collected to get inspiration on which Q\&A we could put in the benchmark. In the following section, we present a selection of two representative user questions and the corresponding responses given by ReservoirChat$_{Big}$ between August and November.

\subsection{Example of knowledge question}

\noindent\textbf{User:}\\
How can a random matrix make such results ? Why does it work ? How can it store such an amount of information ?

\noindent\textbf{ReservoirChat:}\\
A random matrix can make such results because it is a matrix that is generated using random values. This randomness can lead to the creation of complex patterns and structures that can capture and store a significant amount of information.

\noindent
The ability of a random matrix to store information is a result of its dimensionality and the distribution of its values. A high-dimensional random matrix can potentially have a large number of unique configurations, which can be used to represent a wide range of data. The distribution of values in the matrix also plays a role, as certain distributions can lead to more efficient packing of information.

\noindent
However, it's important to note that the effectiveness of a random matrix in storing information can vary depending on the specific characteristics of the data and the task at hand. While random matrices can be powerful tools, they may not always be the best choice for every situation.

\subsection{Example of code question}

\noindent\textbf{User:}\\
Code me the initialization of a reservoir

\noindent\textbf{ReservoirChat:}\\
Sure, I can help you with that. 
 
 \begin{figure}[h]
\begin{lstlisting}
from reservoirpy.nodes import Reservoir, Ridge

# Define the reservoir size and input dimension
reservoir_size = 100
input_dimension = 1

# Initialize the reservoir with random weights
reservoir = Reservoir(reservoir_size, input_dimension, 42)

# Initialize the readout with Ridge regression
readout = Ridge(ridge=1e-5)
\end{lstlisting}

\caption{Simple code produced by \emph{ReservoirChat Big} for the prompt \enquote{Code me the initialization of a reservoir}.}
\centering
\label{fig-code-generated-reservoirchat-big}
\end{figure}
 
In Figure~\ref{fig-code-generated-reservoirchat-big}, on can see a simple example of how you might initialize a reservoir using the ReservoirPy library in Python. This example uses random initialization.
In this code, we first import the necessary classes from the ReservoirPy library. We then define the size of the reservoir and the dimension of the input. Next, we initialize the reservoir with random weights using the `Reservoir` class. Finally, we initialize the readout with Ridge regression using the `Ridge` class.

\section{Discussion}

To summarize, there is a clear need for a tool capable of assisting users working with Reservoir Computing, an underrepresented technology in the literature. While traditional LLMs can partially fulfill this task, they tend to hallucinate and provide incorrect results on their own, especially in specialized areas such as Reservoir Computing. Moreover, even though the ReservoirPy library is acknowledged by the community, there is not as much code available online as for other mainstream libraries such as scikit-learn. Thus, general LLMs do not have enough examples to train on for domain-specific libraries.

We had to make choices during tool development, beginning with Microsoft's GraphRAG library, which was not originally designed to be used with LLMs other than ChatGPT. As a result, we had to identify LLMs and embeddings that match our project. This included requirements such as code generation, answering questions about RCs, available computational power, and support for European languages. 
In this context, Codestral emerged as a suitable choice \cite{jiang2023mistral} \cite{jiang2024mixtral}.
However, running these models locally requires high computing power, and even though Codestral can be considered as a small model with its 22 billions parameters, it still required the use of a GPU (NVIDIA V100 16Go) \cite{bendi2024deploying}. 

ReservoirChat has shown better results on the benchmark in both knowledge-based and coding tasks compared to Codestral, the model on which it is based. Moreover, a larger database intuitively tends to provide an acceptable trade-off between reliability in answering knowledge-based and code-related questions. Additionally, methods implemented in ReservoirChat$_{Big}$ do not produce aberrant ouputs, nor responses entirely different from those of other models. Finally, ReservoirChat$_{Big}$ fulfills its objective, by being able to relevantly answer both knowledge and coding questions towards Reservoir Computing and ReservoirPy, with higher reliability than other tested commonly used LLMs.

It is important to note that GraphRAG was originally developed to extract concepts and relationships from narrative texts. However, the concepts involved in the challenges we encountered, such as code related tasks or technical knowledge in Reservoir Computing, are significantly different and may require prompts tailored to these domains, rather than those originally designed for GraphRAG. A potential improvement could be to refine GraphRAG's prompts to better extract the concepts and relationships relevant to our areas of interest. 

This approach could serve as a model for making technical documentation more interactive.
It could represents a significant advancement in how users engage with specialized libraries and apply them in their own projects. With further refinements, ReservoirChat could become a reliable and practical solution for every Reservoir Computing developer. Fine-tuning hyperparameters and system settings could help optimize response time and quality. Furthermore, careful selection of the LLM remains a critical component of system performance.

Additional testing should also be considered to ensure safety, no inappropriate response should be generated even in unanticipated use cases. Data fairness should be verified, as always, to avoid any misuse of intellectual property, as it is an important concern for LLM pretraining. The knowledge graph method could also be compared to alternative approaches in terms of answer interpretability, server resource requirements, and cost of minimal configuration for acceptable user experience.

The ReservoirChat tool, based on a GraphRAG approach combined with a specialized knowledge base, addresses the practical needs of the Reservoir Computing community. It currently simplifies the learning and application of libraries like ReservoirPy, notably by assisting with debugging and suggesting relevant hyperparameters (e.g., spectral radius, leak rate). For instance, a Phd student tackling their first Echo State Network project can turn to ReservoirChat as a technical partner for configuring and optimizing their network, while an engineer seeking to deploy Reservoir Computing modules in a production environment can quickly resolve compatibility and performance issues.

Moving forward, continuously adding new scientific publications to the knowledge graph will allow ReservoirChat to stay aligned with advances in the field (e.g., hybrid architectures, photonic or spintronic physical reservoirs, and Deep RC). This would enable researchers to interactively consult recent papers and code prototypes to identify promising configurations more efficiently. Moreover, the graph-based approach offers improved traceability and interpretability: by retracing inference paths, users can not only obtain more reliable responses but also deepen their understanding of key Reservoir Computing concepts. These developments position ReservoirChat as a versatile tool that supports research, fosters innovation, and serves as an interactive documentation and educational resource for the Reservoir Computing community.

\bibliographystyle{ieeetran}
\bibliography{references}

\newpage
\section*{\textbf{Acknowledgments}}
We thank the following grants: the Inria AEx BrainGPT project (\url{https://www.inria.fr/en/braingpt}) and the ANR DeepPool ANR-21-CE23-0009 project (\url{https://anr.fr/Projet-ANR-21-CE23-0009}). 
\\
Experiments presented in this paper were carried out using the PlaFRIM experimental testbed, supported by Inria, CNRS (LABRI and IMB), Université de Bordeaux, Bordeaux INP and Conseil Régional d’Aquitaine (see \url{https://www.plafrim.fr}). 
% pour trouver les remerciements de plafrim : https://plafrim-users.gitlabpages.inria.fr/doc/#misc_publications

\vspace{0.5cm}

\newpage

\section*{\textbf{Appendix}}

\subsection*{Pearson Correlation between models for the whole questions}

\begin{figure}[h]
\includegraphics[scale=0.45]{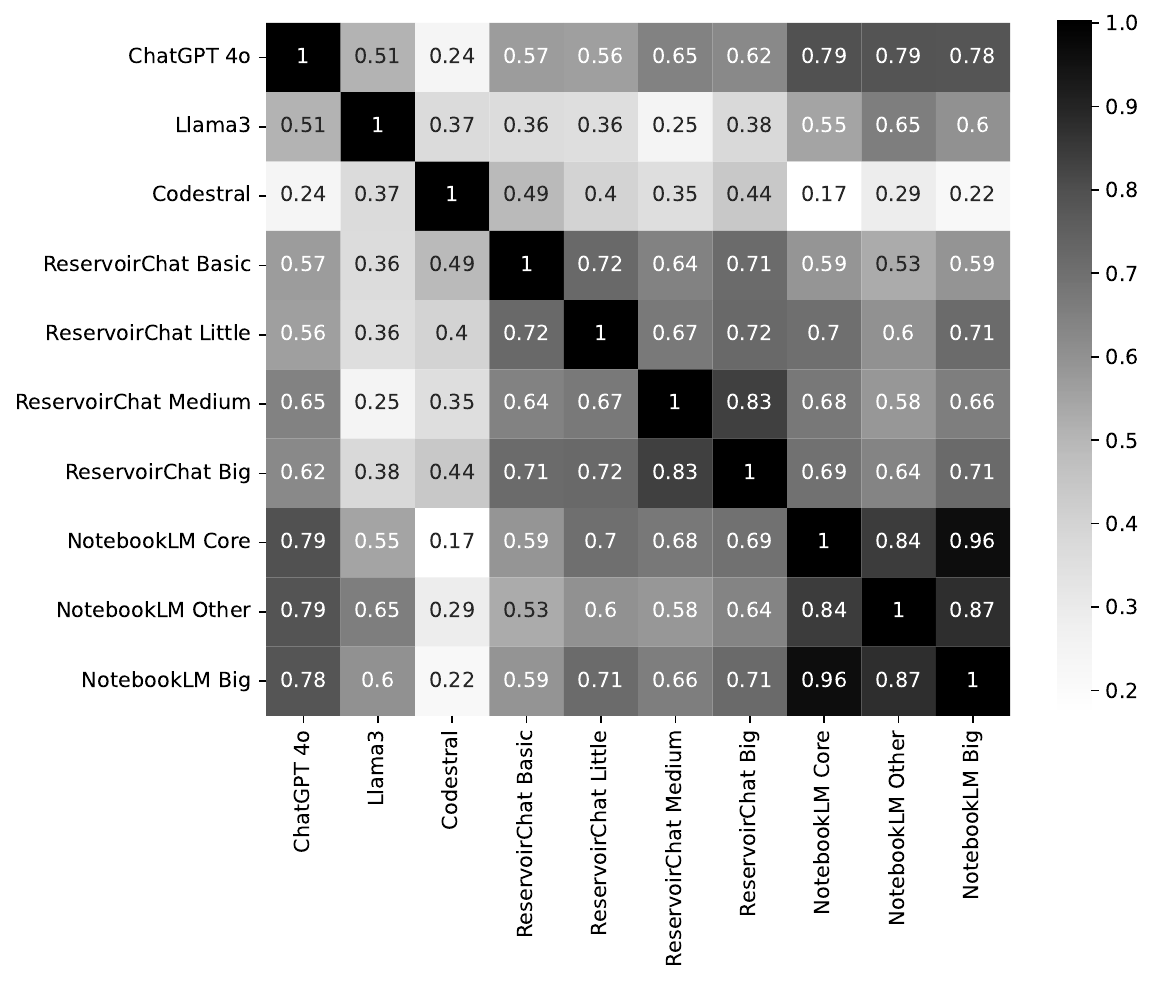}
\centering
\caption{Pearson correlation matrix of every model for the whole questions}
\label{fig-corr-matrix-whole}
\end{figure}

For each model, we concatenated the 3 answers they made for all questions of the benchmark into one vector, and we computed the Pearson correlation between all of them, giving us the Heatmat of correlation that we can see in Figure~\ref{fig-corr-matrix-whole}.

\subsection*{Similarity Rate}

\begin{figure}[h]
\caption{Similarity Rate Table between the answers of ReservoirChat$_{Big}$ and ChatGPT-4o, Llama3, Codestral and NotebookLM$_{Big}$}
\includegraphics[scale=0.75]{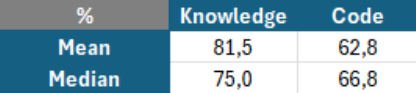}
\centering
\end{figure}

Big ReservoirChat outputs answers significantly similar to the other models, with a correlation rate ranging from 62.8\% to 81.5\%, with coherent medians, not generating any remarkable aberrant error that no other models generated. Therefore, Even though, it has overall best performance than the other models, it seems to give the same good answers and errors than the others. For each question, the mean of the 4 other models answers were divided by the ReservoirChat$_{Big}$ score. Then, each value were put into a series to calculate its mean and median.

\newpage

\section*{\textbf{ReservoirChat: Q\&A benchmark}}

%\xh{mettre en gras plutot que subsubsection}
% C'est bon normalement

\emph{Important: there is only one unique correct answer for each question.}

\subsection*{\textbf{Knowledge questions (Beginner)}}
%\\
%\subsubsection*{\textbf{Beginner}}

\subsection*{1. What is an echo state network?}

\begin{itemize}
  \item A. A place where anything is kept in store.
  \item B. A recurrent neural network in which usually only the output neurons are trained.
  \item C.  A black box component that receives an input signal to be read out and mapped by another process.
  \item D. A large natural or artificial lake used as a source of water supply.
\end{itemize}

Correct: B

\subsection*{2. What is Reservoir Computing?}

\begin{itemize}
  \item A. A machine learning paradigm for timeseries.
  \item B. The construction of a computer model of a petroleum reservoir.
  \item C. A branch of fluid dynamics that studies the design of tank trucks to reduce rollover accidents.
  \item D. It is only a computational framework in which all the computations are done by a pool of water.
\end{itemize}

\subsection*{3. What is call the readout? (in the context of Reservoir Computing)}

\begin{itemize}
  \item A. A linear regression that models the Markovian process of a support vector machine.
  \item B. An efficient way for input processing in order to read only part of the inputs.
  \item C. A black box component that receives an input signal to be read out and mapped back to the input for predictive coding.
  \item D. A trainable layer that extracts the dynamics of a reservoir to produce some desired outputs.
\end{itemize}

Correct: D

\subsection*{4. On which task is Reservoir Computing known to compete with the state-of-the-art?}

\begin{itemize}
  \item A. Video segmentation.
  \item B. Natural language processing.
  \item C. Image recognition.
  \item D. Chaotic timeseries prediction.
\end{itemize}

Correct: D

\subsection*{5. On which task is it challenging to apply Reservoir Computing compared to other state-of-the-art methods?}

\begin{itemize}
  \item A. A task with small univariate dataset.
  \item B. Video segmentation.
  \item C. Chaotic timeseries prediction.
  \item D. Classification of timeseries with few classes.
\end{itemize}

Correct: B

\subsection*{6. Approximately how many neurons are used inside an echo state network?}

\begin{itemize}
  \item A. Around 1 thousand neurons.
  \item B. Around 100 thousand neurons.
  \item C. Around 10 neurons.
  \item D. Around 1 million neurons.
\end{itemize}

Correct: A

\subsection*{7. What is the purpose of using ridge regression instead of linear regression for the readout?}

\begin{itemize}
  \item A. Reduce the computational cost.
  \item B. Improve numerical stability.
  \item C. Improve explainability of the model.
  \item D. Avoid the exploding/vanishing gradient problem.
\end{itemize}

Correct: B

\subsection*{8. How are the weights most often initialized in an echo state network?}

\begin{itemize}
  \item A. They can be randomly initialized and then scaled to have a specific spectral radius.
  \item B. They are tuned according to an autocorrelation Hebbian learning rule.
  \item C. Trained using a linear regression.
  \item D. Each neuron is connected to only one input.
\end{itemize}

Correct: A

%\\
%\subsubsection*{\textbf{Intermediate}}
\subsection*{\textbf{Knowledge questions (Intermediate)}}

\subsection*{9. What is the difference between ‘echo state network’ and ‘Reservoir Computing’?}

\begin{itemize}
  \item A. Reservoir Computing is a type of recurrent neural network model based on the philosophy of the echo state network.
  \item B. There is no difference, we can use the terms Reservoir Computing and echo state network indistinctly.
  \item C. An echo state network is a type of recurrent neural network model based on the Reservoir Computing paradigm.
  \item D. In Reservoir Computing, the reservoir part is a physical system, in echo state networks, the reservoir part is a recurrent neural network.
\end{itemize}

Correct: C

\subsection*{10. Are there other forms of Reservoir Computing than echo state networks?}

\begin{itemize}
  \item A. No, an echo state network is not even a form of Reservoir Computing.
  \item B. Yes, any random kernel method, even those who don't apply to timeseries, are considered to be a form of Reservoir Computing.
  \item C. No, Reservoir Computing only refers to the philosophy behind echo state networks.
  \item D. Yes, there are for instance physical reservoirs or liquid state machines, in which the reservoir is a spiking neural network.
\end{itemize}

Correct: D

\subsection*{11. What is a liquid state machine?}

\begin{itemize}
  \item A. An architecture in which the reservoir part is a pool of spiking neurons.
  \item B. A physical reservoir in which the reservoir is a reservoir of liquid, usually water.
  \item C. Liquid state machine and Reservoir Computing designate the same concept.
  \item D. A computational model of the hippocampus using the Reservoir Computing paradigm.
\end{itemize}

Correct: A

\subsection*{12. Why is it called ‘computing at the edge of chaos’?}

\begin{itemize}
  \item A. Because it is common to add noise (and thus chaos) to the reservoir in order to stabilize its activity.
  \item B. Because Reservoir Computing works best for chaotic timeseries forecasting.
  \item C. Because Reservoir Computing often works best when the dynamics of the reservoir are approaching chaotic dynamics, but are not chaotic.
  \item D. It requires computers to be physically placed at the edge of a chaotic environment, such as a volcano or a kindergarten.
\end{itemize}

Correct: C

\subsection*{13. What is the ‘echo state property’?}

\begin{itemize}
  \item A. The echo state property allows the reservoir to operate with chaotic behavior to enhance computational power.
  \item B. The ability to memorize past activity and cycle over it (the echo).
  \item C. The ability to perfectly reconstruct any input signal from the reservoir activity.
  \item D. The influence of initial states on the reservoir dynamics decays over time, ensuring that the current state primarily reflects recent inputs.
\end{itemize}

Correct: D

\subsection*{14. What are some of the most important hyper-parameters?}

\begin{itemize}
  \item A. The spectral radius of the recurrent weight matrix, the leak rate, the input scaling.
  \item B. The spectral radius of the recurrent weight matrix, the reservoir connectivity, the weight distribution.
  \item C. The spectral radius of the recurrent weight matrix, the reservoir connectivity, the encoding of the input.
  \item D. The activation function, the reservoir connectivity, the regularization parameter.
\end{itemize}

Correct: A

%\\
%\subsubsection*{\textbf{Advanced}}
\subsection*{\textbf{Knowledge questions (Advanced)}}

\subsection*{15. How explainable are Reservoir Computing models?}

\begin{itemize}
  \item A. Reservoir Computing models are generally less explainable due to their reliance on complex, nonlinear dynamics within the reservoir, making it difficult to trace the exact path of information processing.
  \item B. Reservoir Computing models are fully explainable because they rely on predefined, static patterns that do not change over time.
  \item C. Reservoir Computing models are highly explainable because they use simple linear transformations that can be easily interpreted.
  \item D. Reservoir Computing models are completely explainable due to their deterministic nature, which allows for perfect traceability of every computation.
\end{itemize}

Correct: A

\subsection*{16. To what extent do the results vary between two differently initialized reservoirs?}

\begin{itemize}
  \item A. The results between two differently initialized reservoirs are completely unpredictable and random, regardless of the input data.
  \item B. The results between two differently initialized reservoirs are always identical because the initialization has no impact on the reservoir's dynamics.
  \item C. The results between two differently initialized reservoirs can vary somewhat, but the overall performance and behavior of the reservoir are generally robust to different initialisations, provided the reservoir is sufficiently large and well-designed.
  \item D. The results between two differently initialized reservoirs vary significantly because the initialization determines the exact sequence of outputs.
\end{itemize}

Correct: C

\subsection*{\textbf{Knowledge questions (Expert)}}

\subsection*{17. What is the effective spectral radius?}

\begin{itemize}
  \item A. The real spectral radius of the matrix W, that is always a bit different from the specified spectral radius.
  \item B. A weighted sum of all eigenvalues norms, that takes into account the distribution of the spectrum.
  \item C. A value that has similar properties to the spectral radius of a matrix, taking into account the full reservoir equation.
  \item D. The norm of the singular values for each neuron. It is a way to evaluate the impact of each neuron on the reservoir.
\end{itemize}

Correct: C

\subsection*{18. What is a deep reservoir?}

\begin{itemize}
  \item A. A deep reservoir is a Reservoir Computing architecture that consists of multiple layers of interconnected reservoirs, allowing for hierarchical processing and the capture of more complex temporal dynamics.
  \item B. An underground gas or petroleum reservoir that cannot be reached using traditional tools and infrastructure.
  \item C. A deep reservoir is a reservoir that employs deep learning techniques, such as backpropagation, to train the weights within the reservoir, enhancing its ability to learn from data.
  \item D. A deep reservoir is a reservoir that uses extremely large and dense weight matrices to store vast amounts of data.
\end{itemize}

Correct: A

\subsection*{19. What is the use of an orthogonal matrix in the reservoir equation?}

\begin{itemize}
  \item A. An orthogonal matrix can be represented in a condensed form, improving matrix multiplication computation time.
  \item B. An orthogonal matrix in the reservoir equation is used to prevent any interaction between neurons, maintaining independence.
  \item C. This is a trick question, there is no point in using an orthogonal matrix.
  \item D. It augments the memory capacity of the reservoir.
\end{itemize}

Correct: D

\subsection*{20. What is a Conceptor?}

\begin{itemize}
  \item A. A conceptor is a mathematical function used to compress the data within a reservoir, reducing its dimensionality for faster processing.
  \item B. A conceptor is a hardware component that accelerates the computation of reservoir dynamics by offloading calculations to a dedicated processor.
  \item C. A derivation of Reservoir Computing which can store and recall patterns.
  \item D. A conceptor is a specialized type of neuron within a reservoir that is designed to store and retrieve specific concepts or patterns.
\end{itemize}

Correct: C

%\subsection*{\textbf{Code questions}}
%\subsubsection*{\textbf{Code (not debug) questions}}
\subsection*{\textbf{Code questions (not debug)}}

\subsection*{1. I want to train my echo state network on multiple timeseries that have different lengths. I have seen in the documentation that you can put a 3D numpy array with shape (timeseries, timesteps, dimensions), but it wouldn't work in my case as the timeseries have different lengths.}

\begin{itemize}
  \item A. There is no way to do that in ReservoirPy as it is most probably not a good idea to train a model with different lengths and it would induce unexpected results.
  \item B. You can pass a list of 2D numpy arrays that represents timeseries. As lists can contain numpy arrays of different shapes, you can specify timeseries with different lengths.
  \item C. You would have to pad every timeseries with zeros and then concatenate them.
  \item D. As NumPy doesn't support sparse arrays, it is best to use xarray for this use-case.
\end{itemize}

Correct: B

\subsection*{2. Make me a reservoir, with 1000 neurons, and with a uniform distribution of weights, and a sparsity of 95\%.}

\begin{itemize}
  \item A. 
\begin{lstlisting}
from reservoirpy as rpy
reservoir = rpy.nodes.Reservoir(neurons=1_000, connectivity=0.05, weights="uniform")
\end{lstlisting}
  \item B. 
\begin{lstlisting}
from reservoirpy as rpy
reservoir = rpy.nodes.Reservoir(units=1_000, sparsity=0.95, W=rpy.mat_gen.uniform)
\end{lstlisting}
  \item C. 
\begin{lstlisting}
from reservoirpy as rpy
reservoir = rpy.Reservoir(units=1_000, rc_connectivity=0.05, distr="uniform")
\end{lstlisting}
  \item D. 
\begin{lstlisting}
from reservoirpy as rpy
reservoir = rpy.nodes.Reservoir(units=1_000, rc_connectivity=0.05, W=rpy.mat_gen.uniform)
\end{lstlisting}
\end{itemize}

Correct: D

\subsection*{3. Create a model in which there are several reservoirs connected in a chain, and a readout at the end.}

\begin{itemize}
  \item A. 
\begin{lstlisting}
from reservoirpy.nodes import Reservoir, Ridge
model = [Reservoir(100, name="1"), Reservoir(100, name="2"), 
         Reservoir(100, name="3"), Reservoir(100, name="4"), 
         Reservoir(100, name="5"), Ridge(ridge=1e-5)]
\end{lstlisting}
  \item B. 
\begin{lstlisting}
from reservoirpy.nodes import Reservoir, Ridge
model = Reservoir(100, name="1") >> Reservoir(100, name="2") 
        >> Reservoir(100, name="3") >> Reservoir(100, name="4") 
        >> Reservoir(100, name="5") >> Ridge(ridge=1e-5)
\end{lstlisting}
  \item C. 
\begin{lstlisting}
from reservoirpy.nodes import Reservoir, Ridge
model = Reservoir(100) > Reservoir(100) > Reservoir(100) 
        > Reservoir(100) > Reservoir(100) > Ridge(ridge=1e-5)
\end{lstlisting}
  \item D. 
\begin{lstlisting}
from reservoirpy.nodes import Reservoir, Ridge
model = Ridge(ridge=1e-5, previous=Reservoir(100, name="5", 
         previous=Reservoir(100, name="4", 
         previous=Reservoir(100, name="3", 
         previous=Reservoir(100, name="2", 
         previous=Reservoir(100, name="1"))))))
\end{lstlisting}
\end{itemize}

Correct: B

\subsection*{4. Write me an echo state network that can efficiently use the many CPU cores my machine has.}

\begin{itemize}
  \item A. 
\begin{lstlisting}
import reservoirpy as rpy
rpy.set_param("backend", "parallel")

from reservoirpy.nodes import ESN
model = ESN(units=100)
model.fit(train_data, train_data)
\end{lstlisting}
  \item B. 
\begin{lstlisting}
from reservoirpy.utils import parallel
from reservoirpy.nodes import ESN
model = ESN(units=100)
with parallel(n_jobs=-1):
    model.fit(train_data, train_data)
\end{lstlisting}
  \item C. 
\begin{lstlisting}
from reservoirpy.nodes import ESN
model = ESN(units=100, workers=-1)
model.fit(train_data, train_data)
\end{lstlisting}
  \item D. ReservoirPy already parallelizes computation by default to ensure the best performance.
\end{itemize}

Correct: C

\subsection*{5. I have a model with several trainable readouts inside as such:}
\begin{itemize}[label= ]
    \item
\begin{lstlisting}
from reservoirpy.nodes import Reservoir, Ridge

model = Reservoir(100, name="R1") >> Ridge(name="readout1")
model >>= Reservoir(100, name="R2") >> Ridge(name="readout2")
model >>= Reservoir(100, name="R3") >> Ridge(name="readout3")
\end{lstlisting}
\end{itemize}

\textit{How can I fit the model, by specifying the Y values to each Ridge node?}

\begin{itemize}
  \item A. It is not possible to do such a thing in ReservoirPy as it wouldn't make sense.
  \item B. You can pass a dict as a y parameter: `model.fit(X, {"readout1": Y1, "readout2": Y2, "readout3": Y3, })`.
  \item C. You would have to fit each part separately before concatenating them.
  \item D. You can specify the node names as parameters and ReservoirPy will dispatch them correctly: `model.fit(X, readout1=Y1, readout2=Y2, readout3=Y3)`.
\end{itemize}

Correct: B

\subsection*{6. I have a NumPy array \texttt{X} of shape \texttt{(timeseries, timesteps, dimensions)} and I want to classify them according to my \texttt{Y} array of shape \texttt{(timeseries, )} which contains numbers from 0 to 9 according to the class the timeseries belongs to. How can I do that in ReservoirPy?}

\begin{itemize}
  \item A.
\begin{lstlisting}
from reservoirpy.nodes import Reservoir, ScikitLearnNode, Ridge
from sklearn.linear_model import RidgeClassifier

Y_ = Y.reshape(-1, 1, 1).repeat(X.shape[1], 1)

model = Reservoir(1000, lr=0.9, sr=1.0) >> ScikitLearnNode(RidgeClassifier, model_hypers=dict(alpha=1e-8))
model.fit(X, Y_)
\end{lstlisting}

  \item B. Reservoir Computing is only a framework for timeseries forecasting, it is not suited for classification.

  \item C.
\begin{lstlisting}
from reservoirpy.nodes import Reservoir, ScikitLearnNode, Ridge
from sklearn.linear_model import RidgeClassifier

model = Reservoir(1000, lr=0.9, sr=1.0) >> RidgeClassifier(alpha=1e-8)
model.fit(X, Y)
\end{lstlisting}

  \item D.
\begin{lstlisting}
from reservoirpy.nodes import Reservoir, ScikitLearnNode, Ridge
from sklearn.linear_model import RidgeClassifier

Y_ = Y.reshape(-1, 1, 1).repeat(X.shape[1], 1)

model = Reservoir(1000, lr=0.9, sr=1.0) >> ScikitLearnNode(RidgeClassifier)
model.fit(X, Y_)
\end{lstlisting}
\end{itemize}

Correct: A

%\subsection*{Code Debug questions}
%\textbf{Code and Implementation Questions}
\subsection*{\textbf{Code questions (debug)}}

\subsection*{7. Here is my code:}

\begin{itemize}[label= ]
    \item
\begin{lstlisting}
from reservoirpy.nodes import Reservoir, Ridge

model = Reservoir(units=200, lr=0.2, sr=1.0) >> Ridge(ridge=1e-4)

for x_series, y_series in zip(X_train, Y_train):
    model.fit(x_series, y_series, warmup=10)

y_pred = model.run(X_test[0])
\end{lstlisting}
\end{itemize}

\textit{Is that correct?}

\begin{itemize}
  \item A. Calling \texttt{.fit} on a model erases the previous trained results. You can instead call \texttt{.fit} once by passing the lists \texttt{X\_train} and \texttt{Y\_train} as parameters.
  \item B. Everything is correct!
  \item C. \texttt{.fit} method is not suited for online training. Use \texttt{.train} instead.
  \item D. Reservoir parameters should be written in full form: \texttt{leak\_rate}, \texttt{spectral\_radius}.
\end{itemize}

Correct: A

\subsection*{8. Here is my code:}

\begin{itemize}[label= ]
    \item
\begin{lstlisting}
from reservoirpy.nodes import Reservoir, Ridge

model = Reservoir(units=200, lr=0.2, sr=1.0, iss=0.2) >> Ridge(ridge=1e-4)

model.fit(X_train, Y_train, warmup=200)
Y_pred = model.run(X_test)
\end{lstlisting}
\end{itemize}

\textit{I have an error. What's wrong?}

\begin{itemize}
  \item A. \texttt{iss} is not a parameter. For scaling the input, the correct parameter is \texttt{scale\_factor}.
  \item B. Reservoir parameters should be written in full form: \texttt{leak\_rate}, \texttt{spectral\_radius}, \texttt{input\_scaling}.
  \item C. You must first create the reservoir and readout nodes, and then connect them, in three separate lines.
  \item D. \texttt{iss} is not a parameter. For scaling the input, the correct parameter is \texttt{input\_scaling}.
\end{itemize}

Correct: D

\subsection*{9. Here is my code:}

\begin{itemize}[label= ]
    \item
\begin{lstlisting}
from reservoirpy.nodes import Reservoir, RLS

model = Reservoir(units=200, lr=0.2, sr=1.0) >> RLS(alpha=1e-4)

for x_series, y_series in zip(X_train, Y_train):
    model.fit(x_series, y_series, warmup=10)

y_pred = model.run(X_test[0])
\end{lstlisting}
\end{itemize}

\textit{I have an error. What's wrong?}

\begin{itemize}
  \item A. The RLS node can only be trained online, but the \texttt{.fit} method is for offline training. You should use \texttt{.train} instead.
  \item B. The model has been trained on a list of timeseries but is run on a single timeseries.
  \item C. There are not enough units inside the reservoir. For this task, having at least 1000 neurons is recommended.
  \item D. Wrong import path: to import the \texttt{Reservoir} node, use \texttt{from reservoirpy.nodes.reservoirs import Reservoir}.
\end{itemize}

Correct: A

\subsection*{10. Here's my code:}

\begin{itemize}[label= ]
    \item
\begin{lstlisting}
from reservoirpy.nodes import Input, Output, Reservoir, Ridge
R1 = Reservoir(100, lr=0.01, sr=1.)
R2 = Reservoir(100, lr=0.03, sr=1.)
R3 = Reservoir(100, lr=0.09, sr=1.)
R4 = Reservoir(100, lr=0.3, sr=1.)
R5 = Reservoir(100, lr=0.9, sr=1.)
R6 = Reservoir(100, lr=0.01, sr=1.)
readout = Ridge(ridge=1e-5, name="readout")

path1, path2 = R1 >> R6, R2 >> R5
path3 = Input(name="input") >> [R1, R2, R3]
path4 = R1 >> R2 >> R3 >> R4 >> R5 >> R6 >> readout >> Output()
model = path1 & path2 & path3 & path4

model.fit({"input": X_train}, {"readout": Y_train}, warmup=10)
model.run({"input": X_test})
\end{lstlisting}
\end{itemize}

\textit{Is that correct?}

\begin{itemize}
  \item A. The \texttt{.fit} and \texttt{.run} methods only take numpy arrays or list of numpy arrays, not dictionaries.
  \item B. Yes, everything is correct!
  \item C. There is a circular connection in the model.
  \item D. \texttt{path2} is not defined.
\end{itemize}

Correct: B

\subsection*{11. Is this the correct usage of the method \texttt{partial\_fit}?}

\begin{itemize}[label= ]
    \item
\begin{lstlisting}
reservoir, readout = Reservoir(100, sr=1), Ridge(ridge=1e-8)

for x, y in zip(X, Y):
    states = reservoir.run(x)
    readout.partial_fit(states, y)

readout.fit()
model = reservoir >> readout
\end{lstlisting}
\end{itemize}

\begin{itemize}
  \item A. By calling the method \texttt{.fit}, the readout forgets its previous training.
  \item B. The created model won't be fitted.
  \item C. While it works, it can be simplified by creating the model and calling \texttt{partial\_fit} on it.
  \item D. Yes, everything is correctly coded.
\end{itemize}

Correct: D

\subsection*{12. Here is my code:}

\begin{itemize}[label= ]
    \item
\begin{lstlisting}
reservoir, readout = Reservoir(100, sr=1), Ridge(ridge=1e-8)
  
model = reservoir >> readout
  
model.fit(X[:800], Y[:800], warmup=10)
  
steps = 1000
results = np.zeros((steps, 1))

last_output = X[800]
for i in range(steps):
    last_output = model(last_output)
    results[i] = last_output
\end{lstlisting}
\end{itemize}

\textit{Is that the best way to have a model that generates new values by looping on itself?}

\begin{itemize}
  \item A. No, you can connect the readout to the reservoir in order to loop the results back as an input after training: \texttt{readout >> reservoir}.
  \item B. No, it won't work as the reservoir has an input dimension of 100 and the \texttt{results} array containing the results only has its feature dimension set to 1.
  \item C. You can call the \texttt{.autoregress(n=1000)} Model method.
  \item D. Yes, this is probably the best solution.
\end{itemize}

Correct: D

\subsection*{13. Here is my code:}

\begin{itemize}[label= ]
    \item
\begin{lstlisting}
weights = np.random.choice([1, -1], p=[0.6, 1 - 0.6], replace=True, size=(200, 200))
reservoir = Reservoir(W=weights, sr=0.9, lr=0.6)
\end{lstlisting}
\end{itemize}

\textit{I created my reservoir this way, but it seems the reservoir has a very chaotic behavior, even though the spectral radius is below 1.}

\begin{itemize}
  \item A. The rule of the spectral radius <1 holds for matrices with a normal distribution, not a Bernoulli one, which explains why you have a chaotic behavior with a spectral radius of only 1.
  \item B. The rule of the spectral radius <1 only holds when there is no leak rate, so that explains why you have a chaotic behavior with a spectral radius of only 1.
  \item C. The parameter \texttt{sr} is only valid when no weight matrix has been specified. If a matrix is already specified, this argument is ignored.
  \item D. The reservoir argument names are incorrect. You should use \texttt{spectral\_radius} and \texttt{leak\_rate}.
\end{itemize}

Correct: C

%\parageraph{14. Here's my code:}
\subsection*{14. Here's my code:}

\begin{itemize}[label= ]
    \item
\begin{lstlisting}
from reservoirpy.nodes import Input, Output, Reservoir, Ridge
R1 = Reservoir(100, lr=0.01, sr=1.)
R2 = Reservoir(100, lr=0.03, sr=1.)
R3 = Reservoir(100, lr=0.09, sr=1.)
R4 = Reservoir(100, lr=0.3, sr=1.)
R5 = Reservoir(100, lr=0.9, sr=1.)
R6 = Reservoir(100, lr=0.01, sr=1.)
readout = Ridge(ridge=1e-5, name="readout")

path1, path2 = R1 >> R6, R2 >> R5
path3 = Input(name="input") >> [R1, R2, R3]
path4 = R1 >> R2 >> R4 >> R3 >> R5 >> R6 >> readout >> Output()
model = path1 & path2 & path3 & path4

model.fit({"input": X_train}, {"readout": Y_train}, warmup=10)
model.run({"input": X_test})
\end{lstlisting}
\end{itemize}

%\paraegraph{Is that correct?}
\textit{Is that correct?}

\begin{itemize}
  \item A. The \texttt{.fit} and \texttt{.run} methods only take numpy arrays or list of numpy arrays, not dictionaries.
  \item B. Yes, everything is correct!
  \item C. There is a circular connection in the model.
  \item D. \texttt{path2} is not defined.
\end{itemize}

Correct: B

%\PB{On devrait plutôt mettre les références avant l'annexe, non ?}

\end{document}